\documentclass[11pt]{article}
\usepackage{graphicx,amssymb,amsmath}
\usepackage{url}
\usepackage[hang]{subfigure}
\usepackage[all]{xy}

\newtheorem{theorem}{Theorem}

\newtheorem{lemma}[theorem]{Lemma}
\newtheorem{cor}[theorem]{Corollary}

\newtheorem{prop}[theorem]{Proposition}
\newtheorem{open}{Open Problem}

\def\QED{\ensuremath{{\square}}}
\def\markatright#1{\leavevmode\unskip\nobreak\quad\hspace*{\fill}{#1}}
\newenvironment{proof}
  {\begin{trivlist}\item[\hskip\labelsep{\bf Proof.}]}
  {\markatright{\QED}\end{trivlist}}

\long\def\ignore#1{\relax}

\def\tomath#1{\relax\ifmmode#1\else$#1$\fi}

\def\fourldots{\mathinner{\ldotp\ldotp\ldotp\ldotp}}
\def\fourdots{\relax\ifmmode
    \fourldots\else$\mathsurround=0pt \fourldots\,$\fi
    \spacefactor=3000}

\def\nopar#1\par{}
\def\slw #1 {{\sl #1 }}
\def\itw #1 {{\it #1 }}
\def\ttw #1 {{\tt #1 }}
\def\scw #1 {{\sc #1 }}
\def\bfw #1 {{\bf #1 }}

\def\calw #1 {\tomath{{\cal #1}} }

\def\ray#1{\hbox{\vbox{\offinterlineskip\setbox0\hbox{$#1$}
    \hbox to \wd0{\hss$\rightharpoonup$\hss}\vskip-1.0pt\box0}}}
\def\lin#1{\hbox{\vbox{\offinterlineskip\setbox0\hbox{$#1$}
    \hbox to \wd0{\hss$\longleftrightarrow$\hss}\vskip-1.0pt\box0}}}

\long\def\comm#1{\ignorespaces}

\def\raggedcenter{\advance\leftskip by 0pt plus 40em\rightskip=\leftskip
   \parfillskip=0pt \spaceskip=.3333em \xspaceskip=.5em
   \pretolerance=9999 \tolerance=9999
   \hyphenpenalty=9999 \exhyphenpenalty=9999 }

\def\rmop#1(#2){\tomath{\mathop{\mathrm{#1}(#2)}}}
\def\op#1(#2){\tomath{\mathop{\mathtt{#1}(#2)}}}
\def\bfop#1(#2){\tomath{\mathop{\mathbf{#1}(#2)}}}
\def\itop#1(#2){\tomath{\mathop{\mathit{#1}(#2)}}}

\def\secA{sec} 
\newcommand{\secLab}[1]{\label{\secA:#1}}
\newcommand{\secRef}[1]{\ref{\secA:#1}}

\def\sbsA{sbs} 
\newcommand{\sbsLab}[1]{\label{\sbsA:#1}}

\def\thmA{thm} 
\newcommand{\thmLab}[1]{\label{\thmA:#1}}
\newcommand{\thmRef}[1]{\ref{\thmA:#1}}

\def\lemA{lem} 
\newcommand{\lemLab}[1]{\label{\lemA:#1}}
\newcommand{\lemRef}[1]{\ref{\lemA:#1}}

\def\corA{cor} 
\newcommand{\corLab}[1]{\label{\corA:#1}}
\newcommand{\corRef}[1]{\ref{\corA:#1}}

\def\propA{prop} 
\newcommand{\propLab}[1]{\label{\propA:#1}}

\def\figA{fig} 
\newcommand{\figLab}[1]{\label{\figA:#1}}
\newcommand{\figRef}[1]{\ref{\figA:#1}}

\def\eqA{eq} 
\newcommand{\eqLab}[1]{\label{\eqA:#1}}

\newenvironment{proofL}{\noindent {\bf Proof of Lemma}}{{\hfill $\Box$}}

\def\qed{$\Box$}

\def\sectionD#1#2{
\section{#1}
\secLab{#2}
}

\def\subsectionD#1#2{
\subsection{#1}
\sbsLab{#2}
}

\def\thmD#1#2{
\begin{theorem}
\thmLab{#1}
#2
\end{theorem}
}

\def\lemmaD#1#2{
\begin{lemma}
\lemLab{#1}
#2
\end{lemma}
}

\def\propD#1#2{
\begin{prop}
\propLab{#1}
#2
\end{prop}
}

\def\corD#1#2{
\begin{cor}
\corLab{#1}
#2
\end{cor}
}

\title{Flattening Single-vertex Origami: the Non-expansive Case}

\author{\setcounter{footnote}{0}%
\def\thefootnote{\arabic{footnote}}
Gaiane Panina$^1$ and
Ileana~Streinu$^2$ 
}

\begin{document}

\date{}
\maketitle


{
\setcounter{footnote}{0}
\def\thefootnote{\arabic{footnote}}
\footnotetext[1]{Institute for Informatics and Automation, V.O. 14 line 39, St.Petersburg 199178,Russia. 
\url{panina@iias.spb.su}
}
\footnotetext[2]{Department of Computer Science, \  Smith College,
Northampton, \ MA 01063, USA. \  \url{istreinu@smith.edu}. 
}
}

\begin{abstract}
A {\em single-vertex origami} is a piece of paper with straight-line rays called creases emanating from a fold vertex placed in its interior or on its boundary. The {\em Single-Vertex Origami Flattening} problem asks whether it is always possible to reconfigure the creased paper from any configuration compatible with the metric, to a flat, non-overlapping position, in such a way that the paper is not torn, stretched and, for {\em rigid origami}, not {\em bent} anywhere except along the given creases.
	
Streinu and Whiteley showed how to reduce the problem to the carpenter's rule problem for spherical polygons. Using spherical expansive motions, they solved the cases of open $< \pi$ and closed $\le 2 \pi$ spherical polygons. Here, we solve the case of open polygons with total length between $[\pi, 2\pi)$, which requires non-expansive motions. Our motion planning algorithm works in a finite number of discrete steps, for which we give precise bounds depending on both the number of links and the angle deficit.
\end{abstract}

\medskip

%
%

\maketitle


\sectionD{Introduction}{introduction}

In this paper, we answer in the affirmative the following conjecture of \cite{streinu:whiteley:origami:2005}:

\medskip
{\em A single-vertex origami whose fold vertex is placed on the boundary of the paper can {\em always} be reconfigured to the flat position with a non-colliding continuous motion.} 

\medskip
\noindent
The technical formulation of the problem is given below.

\medskip
\noindent
{\bf Rigid Origami.} An origami is a flat piece of paper marked with a straight-line plane graph drawing. Fig.~\figRef{flatOrigami} exemplifies the first non-trivial type, which is the topic of this paper: an origami with just a single vertex. By creasing the paper along the edges and possibly bending the paper while maintaining its intrinsic metric (i.e. not allowing any tearing or stretching), the origami will take various 3D shapes. {\em Rigid origami} is the study of those configurations and motions which further restrict the faces to remain planar. Thus they behave like rigid panels hinged along the crease lines, along which they may rotate. While in practice the paper bends during folding, this model offers a rigorous mathematical formulation and potential for algorithmic treatment.

In the quest of mathematical laws for origami folding, we start with the simplest situation: the {\bf single-vertex origami}. This is the case of a single vertex,  with edges emanating from it that partition the paper into wedges, as in Fig.~\figRef{singleVertexInterior}.

\medskip

\begin{figure}[h]
    \centering
    \subfigure[]{\figLab{singleVertexInterior}\includegraphics[width=0.25\textwidth]{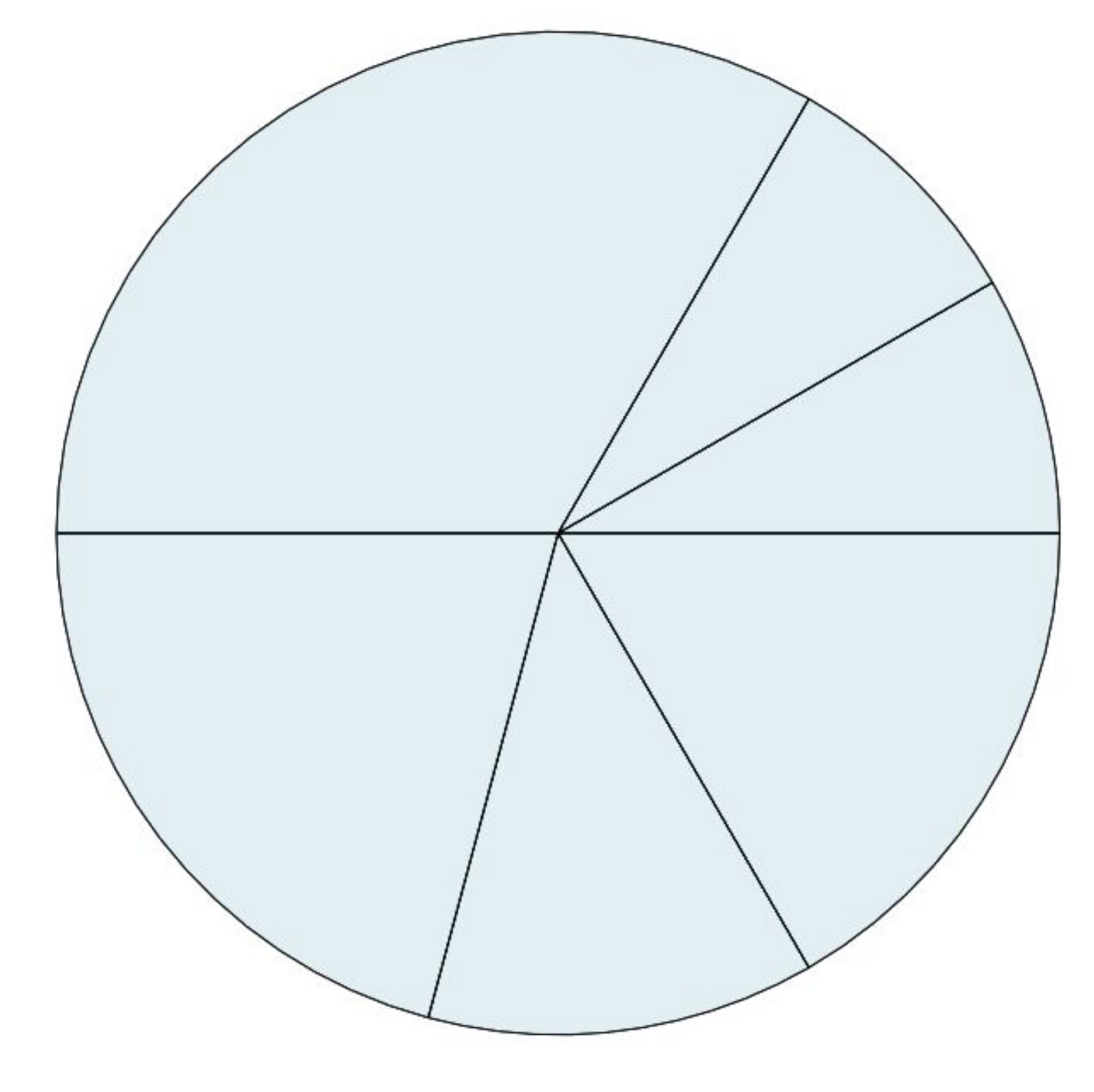}}
    \subfigure[]{\figLab{singleVertexFolded}\includegraphics[width=0.25\textwidth]{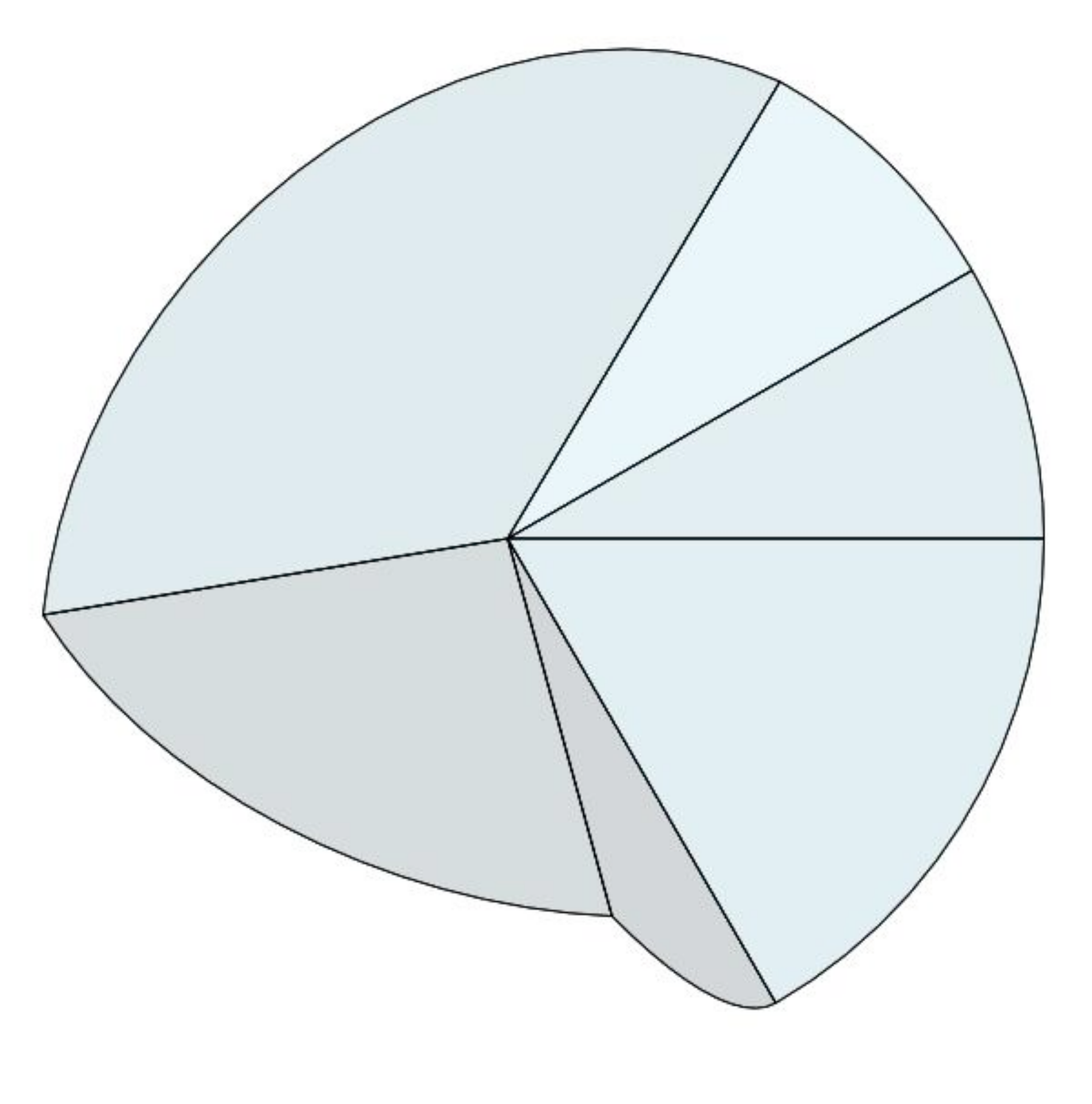}}
    \subfigure[]{\includegraphics[width=0.25\textwidth]{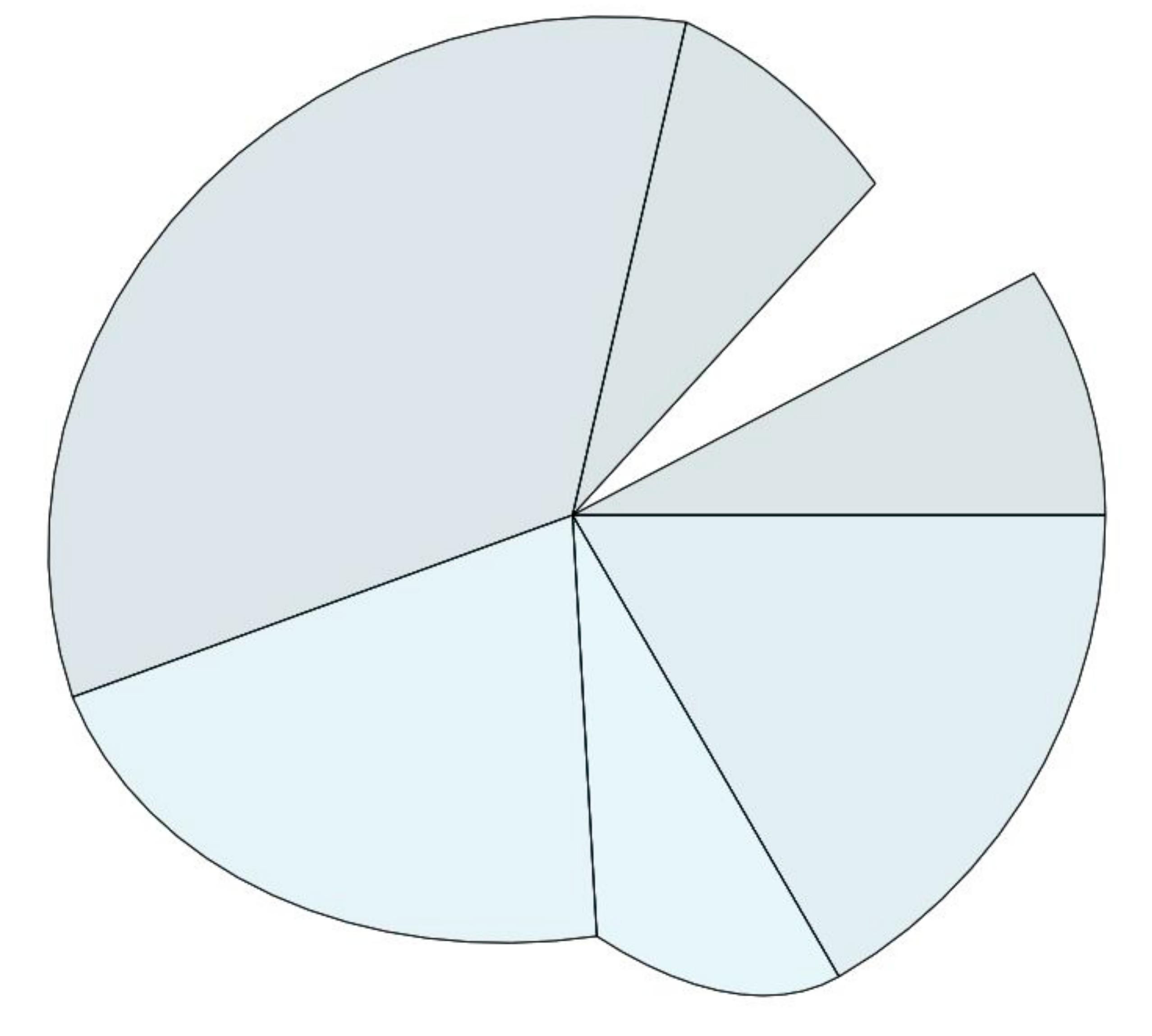}}
    \caption{(a) Single vertex origami with interior fold vertex, in a flat non-overlapping configuration and (b) in a 3D folded configuration. For contrast, if a crease in the origami from (a) is cut up, then the fold vertex is no longer interior; a 3D folded shape of this case is illustrated in (c).}
    \figLab{flatOrigami}
\end{figure}

\noindent
{\bf Non-Flat Paper Folding.} It is worth mentioning that the abstract single-vertex origami problem includes {\em non-flat paper}, looking very much like the corner of a polyhedral surface, rather than a flat sheet. The angle sum around the vertex may be smaller than $2\pi$, as in the case of a convex polyhedron vertex, or larger than $2\pi$, as in saddle surfaces and hyperbolic virtual polytopes \cite{panina:virtualPolytopesPseudoTriangulations:2006}. It equals $2\pi$ for an interior-vertex origami as in Figures~\figRef{singleVertexInterior},~\figRef{singleVertexFolded}.

\medskip
\noindent
{\bf Folding rigid origami.} 
Very little is known mathematically about rigid origami. According to T. Hull's web page \cite{hull:rigidOrigami:2003} devoted to the topic, as of 2003 only two papers have been published in this area.  The classical origami literature is concerned mostly with characterizations of folded states and axiomatics for folding patterns. 
Equally interesting and important, but also very little studied is the motion planning problem for origami, i.e. the design of reconfiguration trajectories. 
In particular, the non-self-intersecting foldability and reconfiguration of rigid origami has received so far {\em very} little attention, mostly because it is a very difficult problem. Section 12.3 of \cite{GFALOP} summarizes in a little over one page what is known about continuous foldability of single-vertex origami, which is essentially the previous paper \cite{streinu:whiteley:origami:2005} of the second author and Whiteley.

Meanwhile, the topic gained momentum due to new robotics applications \cite{lu:akella:foldingCartons:2000,balkom:mason:origamiFolding:2004} and the advent of practical nano-origami at the DNA \cite{rothemund:foldingDNA:2006} and mechanical \cite{Jurga:Hidrovo:Nanostructured:2003,in:kumar:shao-horn:barbastathis:Origami2006,schmitt:streinu:cognitive:2004} level. In simulations, the probabilistic roadmap algorithm has been shown to find folding trajectories for small "paper craft" puzzles \cite{song:amato:motionPlanningPaperCraft:2001}. Understanding the laws of origami {\em folding} is becoming a recognized area of mathematical and algorithmic research, as the field itself moves from recreational aspects to increasingly practical applications.

\medskip
\noindent
{\bf Single-vertex Origami.} One of the main questions in algorithmic origami is: {\em Are there origami folded shapes which are compatible with the creases and the induced metric of the paper, but which cannot be folded by a collision-avoiding motion? } Here we show that there aren't any single-vertex ones. Moreover, the reconfiguration of a single-vertex origami between two configurations can be performed algorithmically, in finite time.

\begin{figure}[h]
    \centering
    \subfigure[]{\figLab{origamiEdge}\includegraphics[width=0.3\textwidth]{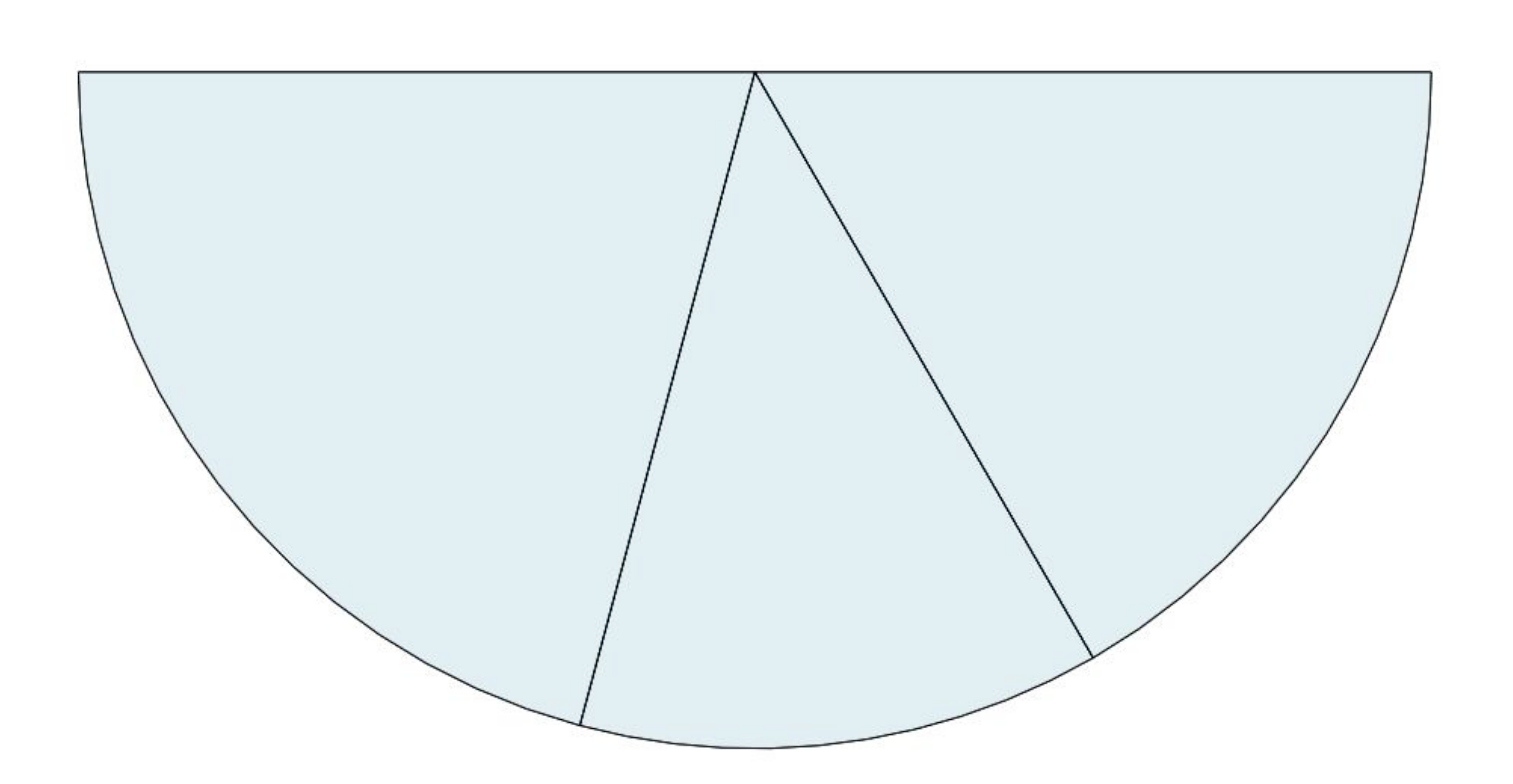}}
    \subfigure[]
	{\figLab{origamiCornerR}
	\includegraphics[width=0.3\textwidth]{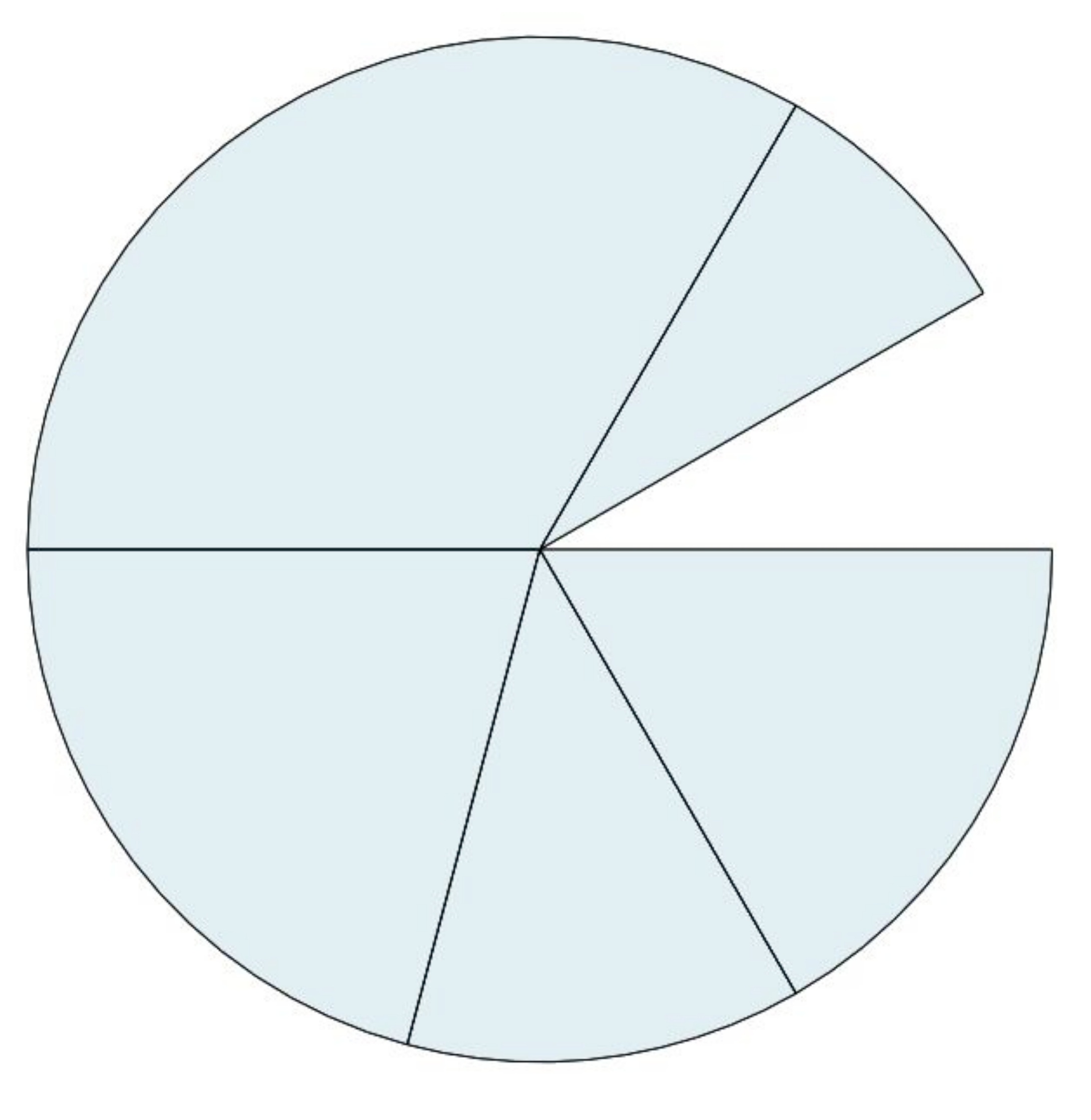}}
    \subfigure[]{\figLab{origamiCornerC}\includegraphics[width=0.3\textwidth]{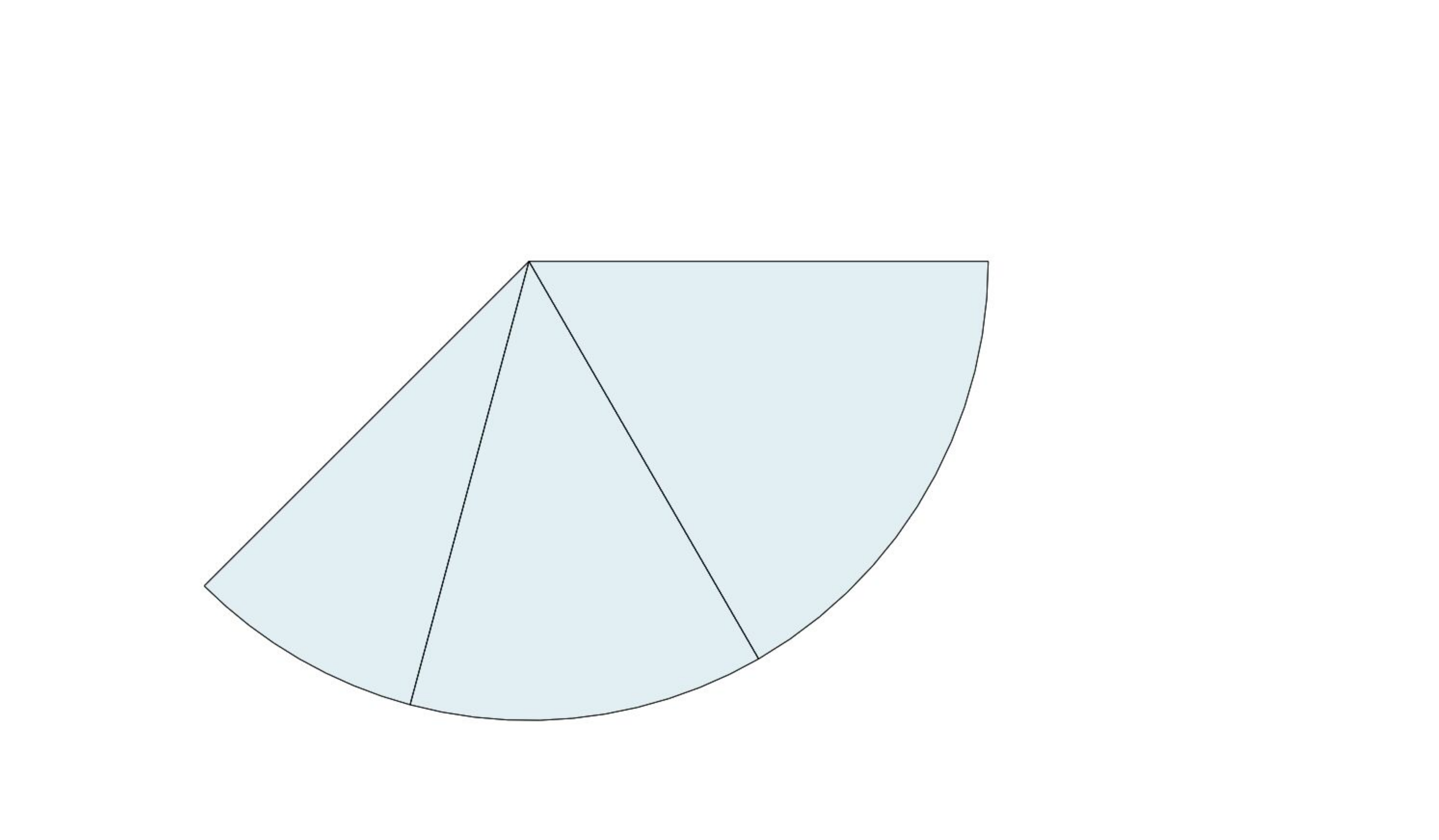}} 
    \caption{Single-vertex origami: the {\em vertex-on-boundary} case. The fold vertex is placed: (a) on an edge of the paper, (b) at a reflex  corner and (c) at a  convex corner.}
    \figLab{svo}
\end{figure}

For a vertex placed in the interior of the piece of paper, as in Fig.~\figRef{singleVertexInterior}, or for a boundary vertex incident to a paper angle\footnote{Only the angles incident to the fold vertex are relevant to the question.} of at most $\pi$, as in Figures~\figRef{origamiEdge} and \figRef{origamiCornerC}, Streinu and Whiteley~\cite{streinu:whiteley:origami:2005} answered the problem in the affirmative by reducing it to a version of the carpenter's rule problem \cite{connelly:demaine:rote:carpenterRule:2003,streinu:pseudoTriang:dcg-focs:2005} extended from planar to spherical polygons. The method used in  ~\cite{streinu:whiteley:origami:2005}, based on expansive motions (in particular those induced by pointed spherical pseudo-triangulation mechanisms),  can be applied to unfold closed spherical polygons  of total length less than $2\pi$, and open ones less than $\pi$ (precise definitions of single-vertex origami angle size and spherical polygon length will be given in Section \secRef{preliminaries}). The algorithm uses at most $O(n^3)$ steps, each one being induced by the well-defined expansive direction of motion of a pseudo-triangulation one-degree-of-freedom mechanism. The paper \cite{streinu:whiteley:origami:2005} also showed that spherical polygons of lengths larger than $2 \pi$ (and, equivalently, single-vertex origamis of total angle larger than $2 \pi$) may not be reconfigurable between any two configurations. 

	The remaining case, of open polygons whose length lies between $\pi$ and $2 \pi$, or single-vertex origami incident to a reflex paper corner as in Fig.~\figRef{origamiCornerR},  is not directly amenable to the expansive motion and the pseudo-triangulation techniques, as it requires both contractive and expansive motions. 
	
In this paper we settle the problem for spherical bar-and-joint polygonal paths of total length $\alpha\in (\pi,2 \pi)$, by showing that it is always possible to unfold them without self-collisions. The motion (necessarily partially non-expansive) can be carried out in discrete steps, and completed in finite time, for which we give precise bounds. However, the bound on the number of steps will depend not just on $n$ (the number of links in the chain), but also on the angle deficit $2 \pi - \alpha$.

\sectionD{Preliminaries}{preliminaries}

We start by introducing the relevant concepts, and we summarize the previous results which reduce the single-vertex origami problem to the spherical carpenter's rule. Then we give a brief summary of the techniques we rely on for establishing the main result of the paper, as well as an overview of concepts from spherical geometry needed in our proofs.

\medskip

\subsectionD{From origami to spherical chains}{origamiToChains}

\medskip
\noindent
{\bf Single-vertex origami.} \ A {\em single vertex origami} \ is a bounded or unbounded piece of paper, together with a point on it (the {\em fold vertex}) and a finite set of rays emanating from the vertex, called {\em creases}. The rays induce a natural  ordering around the vertex. If the vertex is placed in the interior of the piece of paper, as in Fig.~\figRef{singleVertexInterior}, then this is the circular counter-clockwise (ccw) ordering of the rays around the vertex. If the vertex is placed on the boundary, as in Fig.~\figRef{origamiCornerR}, then the ordering is linear and starts and ends at an edge along the paper's boundary. These extreme edges are not creases.

\medskip
\noindent
{\bf Single-vertex panel-and-hinge chains.}
We consider only {\em rigid origami}, where the wedge-like flat regions between two consecutive creases maintain their intrinsic and extrinsic metric, i.e. they behave like rigid flat metal {\em panels} rather than flexible paper. Single-vertex origami can now be modeled as a collection of rigid polygons (the panels) connected by hinges (the creases), such that all the hinges are concurrent in a single vertex.

For single-vertex origami, the shape of the polygonal panels is irrelevant to questions of self-intersection. They may even be unbounded. All that matters is where the vertex is placed: in the interior of the paper, or on its boundary. In the first case, the panel-and-hinge structure forms a {\em closed} chain. In the second, it is an {\em open chain}.

\begin{figure}[htp]
  \begin{center}
    \subfigure[]
	{\figLab{svOrigami}
	\includegraphics[width=0.3\textwidth]{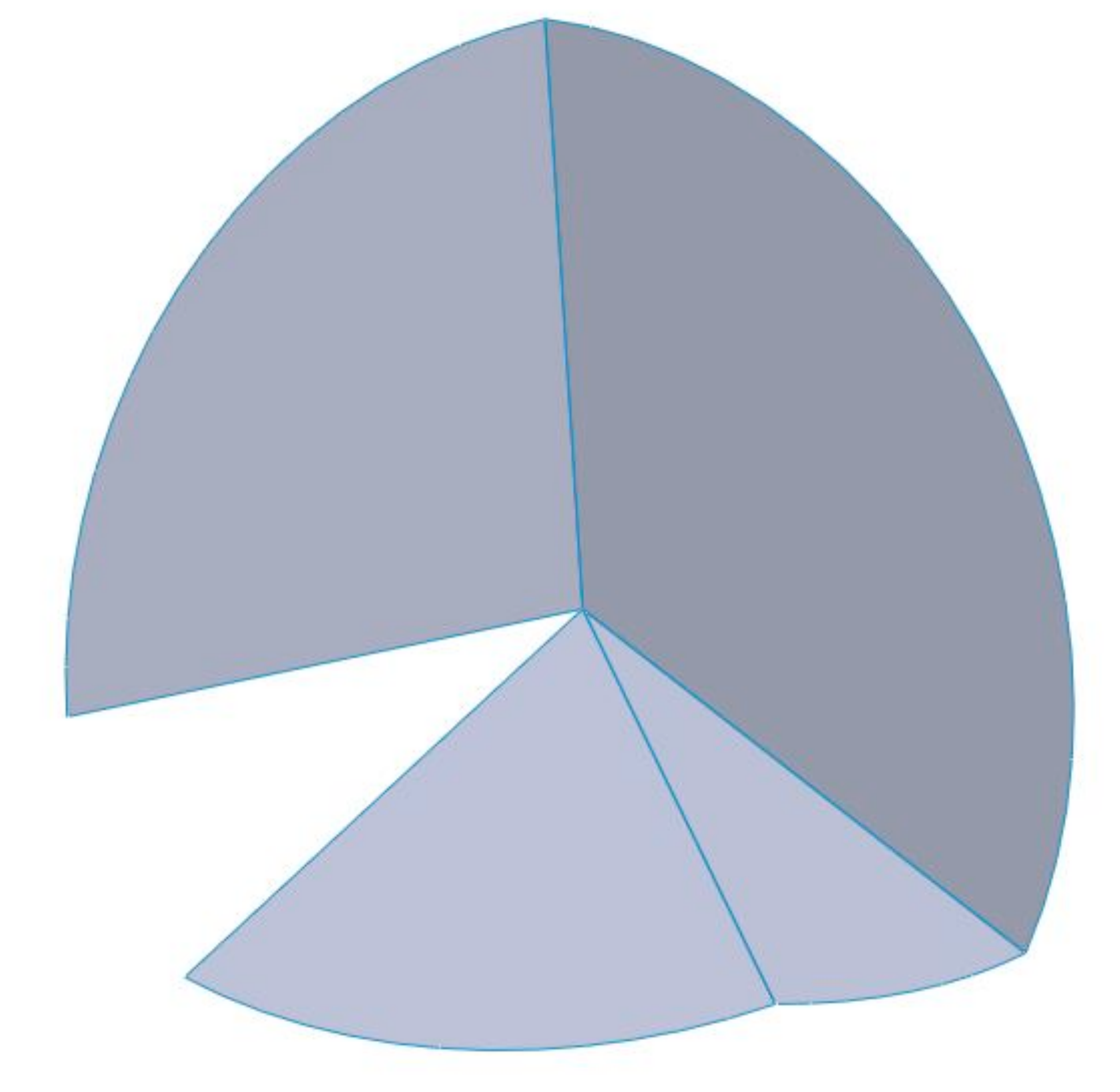}}
    \subfigure[]
	{\figLab{chainOfArcs}
	\includegraphics[width=0.3\textwidth]{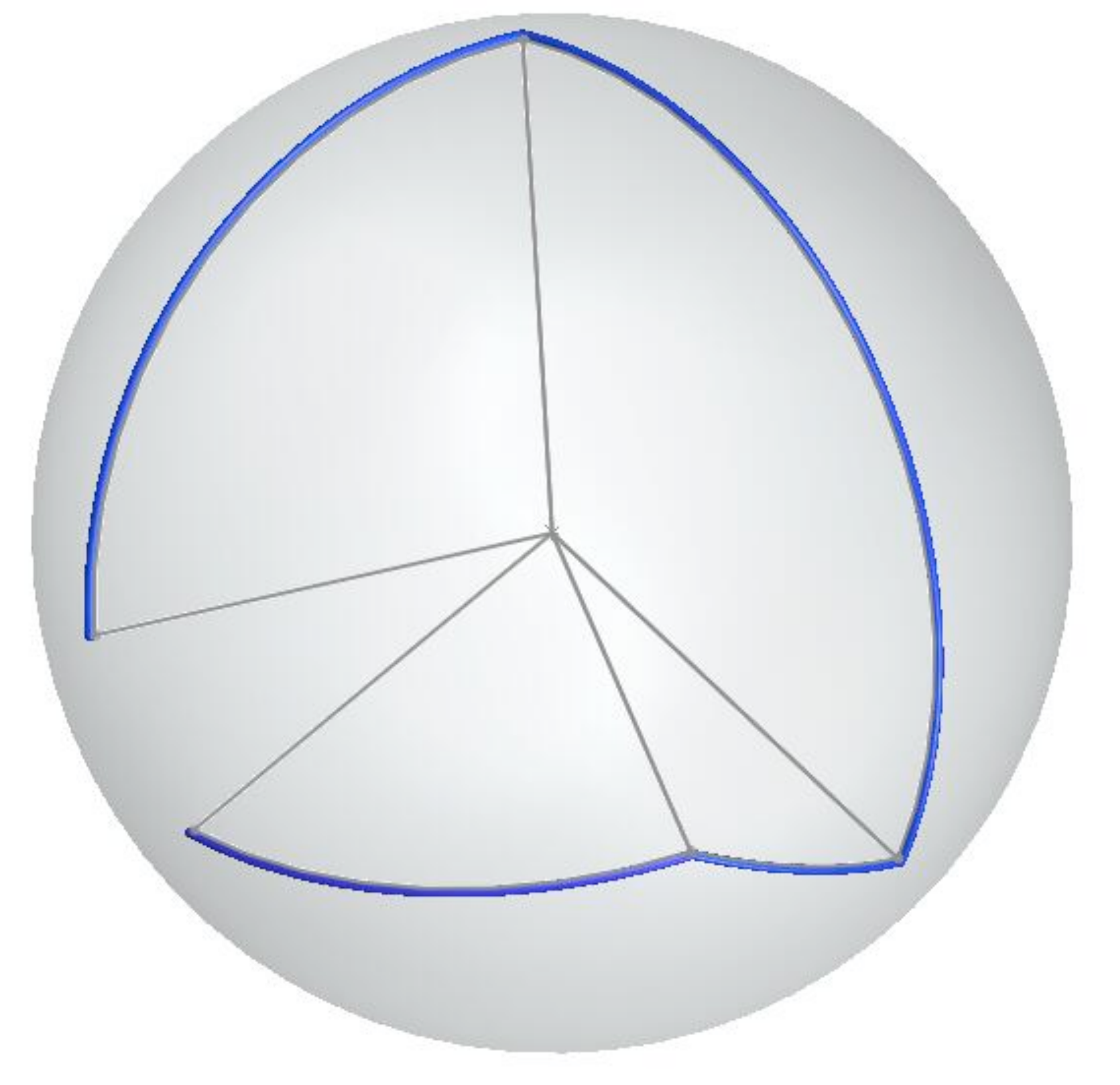}} 
  \end{center}
  \caption{Two views of an open single vertex origami, in a 3D folded position: (a) as a panel-and-hinge chain, and (b) as a spherical polygonal chain.}
  \figLab{twoViews}
\end{figure}

\medskip
\noindent
{\bf Spherical chains and polygons.}
We cut a circle (of sufficiently small radius so that it crosses all the crease lines), centered at the vertex. From now on, we work with this bounded piece of paper, as in Figures~\figRef{flatOrigami} and \figRef{svo}. 
Each panel is now bounded by two straight line edges (corresponding to either the creases, or to the paper boundary) and a circular arc. In any configuration of the origami in three-dimensional space (3D), these circular arcs are arcs of great-circles on a sphere (assumed to be the unit sphere) centered at the fold vertex. 

When the fold vertex is on the boundary of the piece of paper and the spatial origami configuration does not bring (or glue) the boundary edges together, the arcs will form a spherical polygonal path or chain. When the vertex is interior to the original piece of paper, or the folding glues boundary edges, then it will be a closed polygon, as in Fig.~\figRef{singleVertexFolded}. 
Notice that the panels intersect if and only if their corresponding circular arcs  intersect on the sphere.

{\em From now on, we work with the spherical polygonal chain model}, and assume we are on the unit sphere.

\medskip
\noindent
{\bf Notation.} A spherical polygonal chain with $n$ edges is given by an ordered set of points on the unit sphere $p=\{p_0, p_1,\cdots,p_n\}$. Its edges are denoted by $e_i=(p_{i-1},p_{i})$. A {\em subchain} $p[i:j]$ consists in all vertices and edges between the points $p_i$ and $p_j$. Closed polygons appear only indirectly in this paper, via references to previous work, so we do not introduce any special notation for them.

\medskip
\noindent
{\bf Arc length.} A spherical arc has a length, which is measured by the angle at the center of the sphere between the two rays that span the arc. A {\em short arc} has length less than $\pi$. A {\em long} one has length larger than $\pi$, and less than $2\pi$. Arcs larger than $2\pi$ are self-overlapping and hence not within the scope of this paper. All throughout, we work only with short arcs.

\medskip
\noindent
{\bf Chain length.} The {\em length} of a spherical polygonal chain is the sum of its arc lengths. We distinguish three categories of chains: {\em short}, of length strictly less than $\pi$, {\em medium} (the case considered in this paper), of length between $[\pi,2\pi)$, and {\em large}, those exceeding $2\pi$.

\medskip
\noindent
{\bf Configuration space.} The set of all the possible positions of the chain vertices which are compatible with the given edge lengths, up to spherical rigid motions (rotations around the center) is called the {\em configuration space} of the chain. To eliminate the rigid motions, we can {\em pin down} any edge.

\medskip
\noindent
{\bf Flat, hemispherical and sphere-spanning chains.}
A chain configuration stretched along a great-circle will be called {\em flat}. If it is contained in some open hemisphere, we call it {\em hemispherical}. Otherwise it is called {\em sphere-spanning}. For example, any closed or open polygonal chain of length at most $\pi$ is hemispherical. When the length exceeds $\pi$, some configurations may be hemispherical, others may not. 

\medskip

\subsectionD{Unfolding spherical chains: previous results}{prevResults}

\medskip
\noindent
{\bf Hemispherical versus planar chains.} 
For points and edges lying in a hemisphere, in particular for a hemispherical chain, oriented matroid concepts such as pointedness and convex hulls are in one-to-one correspondence with their planar counter-parts. This allows us, among others, to define spherical pointed pseudo-triangulations and apply all the results on polygon unfolding from \cite{streinu:pseudoTriang:dcg-focs:2005}. We refer the reader to that paper,  or to the survey \cite{streinu:rote:santos:survey:2008} for background material on pointed pseudo-triangulations. These concepts are not needed here, because we will use Streinu's pseudo-triangulation-based unfolding algorithm \cite{streinu:pseudoTriang:dcg-focs:2005} only as a black box.

\medskip
\noindent
{\bf Unfolding single-vertex origami: summary of previous results.} There are two types: (a) {\em long chains} (open or closed) may {\em not} be reconfigurable, and (b) {\em short chains} and {\em medium closed chains} can always be reconfigured. For closed chains, the reconfiguration is carried out in the same orientation class. The main idea is that short chains and medium polygons are confined to a hemisphere. In this case, a theorem of \cite{streinu:whiteley:origami:2005} establishes an equivalence between infinitesimal motions of hemispherical and planar polygons, which transfers all the results of the planar carpenter's rule problem to the spherical setting. This equivalence is at the infinitesimal level, and holds whenever a chain configuration can be confined to some hemisphere. It does not require that the total length of the chain be short. We will make substantial use of this observation.

\medskip
\noindent
{\bf When is a chain hemispherical?} \ 
Short chains and \ medium polygons are {\em always} confined to a hemisphere:  this property seems obvious, but its formal proof has not appeared before. It is a simple consequence of our Separation Theorem (theorem ~\thmRef{separation} in Section \ref{sec:proofs}). In the planar carpenter's rule problem, an open chain was treated by just closing it with additional edges. On the sphere, we cannot add edges without increasing the total length.  

In all the other cases, some chain configuration {\em may} span the sphere. If it does not, the motion can proceed, expansively, until the chain touches a great-circle in at least three points. 

The crux of our argument is the treatment of the case of a medium-length chain in a configuration which spans the sphere. The following classical concepts will be needed.

\medskip

\subsectionD{Spherical polar-duality}{polarDuality}

\medskip
Computational Geometers are familiar with planar dualities between points and lines, and their incidence and orientation preserving properties. Here, we make use of their spherical counterpart, which has even stronger, measure-theoretic properties on which we rely in our proofs. 
\begin{figure}[h]
    \centering
    \includegraphics[width=0.3\textwidth]{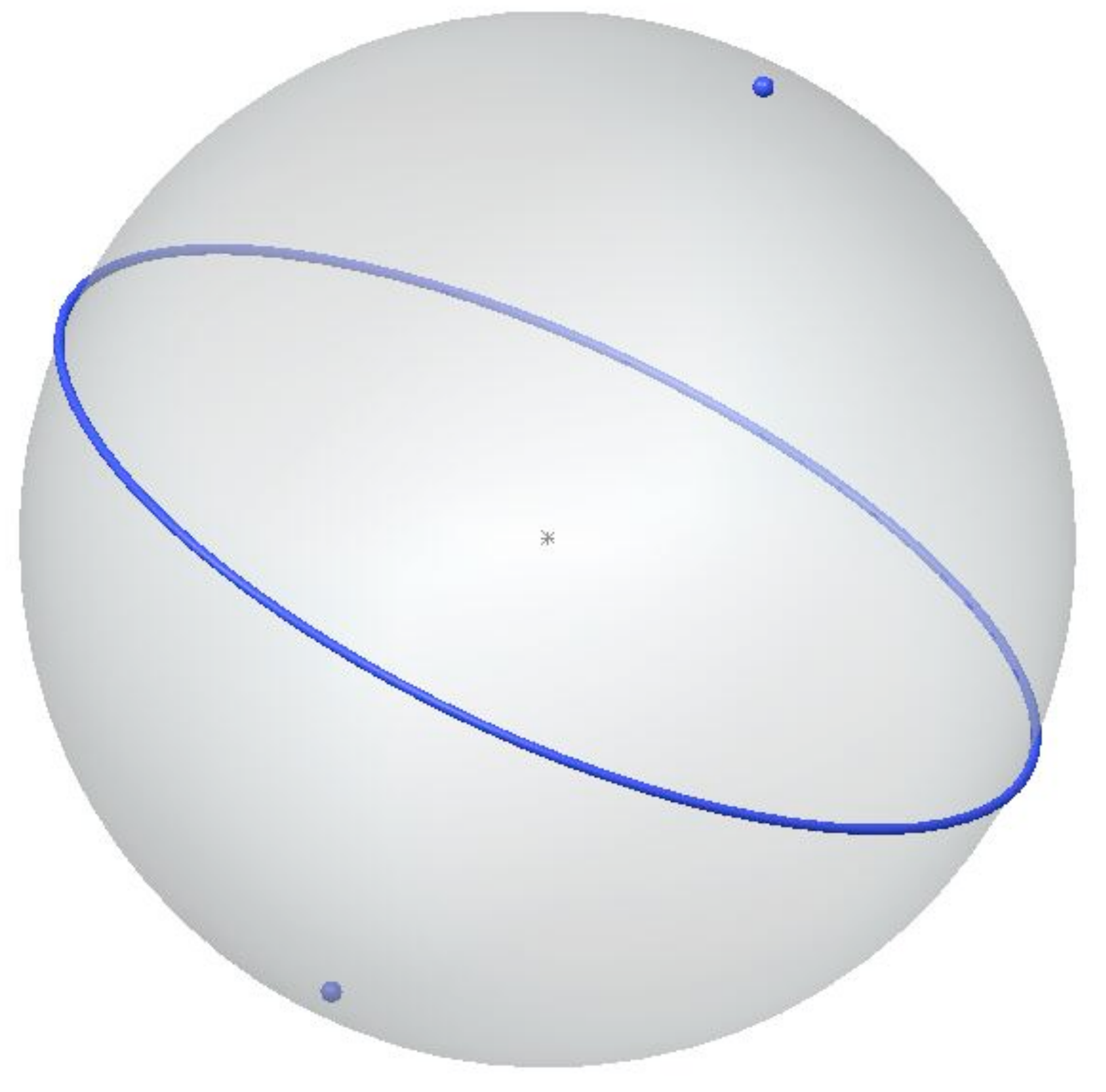}
    \caption{The point-to-great-circle duality on the sphere.}
    \figLab{duality}
\end{figure}

\medskip
\noindent
{\bf Spherical polar-duality.}
The well-known duality between great-circles and antipodal pairs of points takes a great-circle $c$ (viewed as an {\em equator}) to a pair of antipodal points called its {\em poles}, as in Fig. \figRef{duality}. The poles are the intersection points of the sphere with the line orthogonal to the supporting plane of the great circle, and going through the center of the sphere.

This duality (usually referred to as a polar-duality) has all the good incidence and orientation-preserving properties of planar dualities familiar to Computational Geometers, and more. In fact, the {\em natural} definition of point-line duality is the spherical version. We will make use of the following properties.

\begin{figure}[h]
	\vspace{0.3 in}
    \centering
    \includegraphics[width=0.3\textwidth]{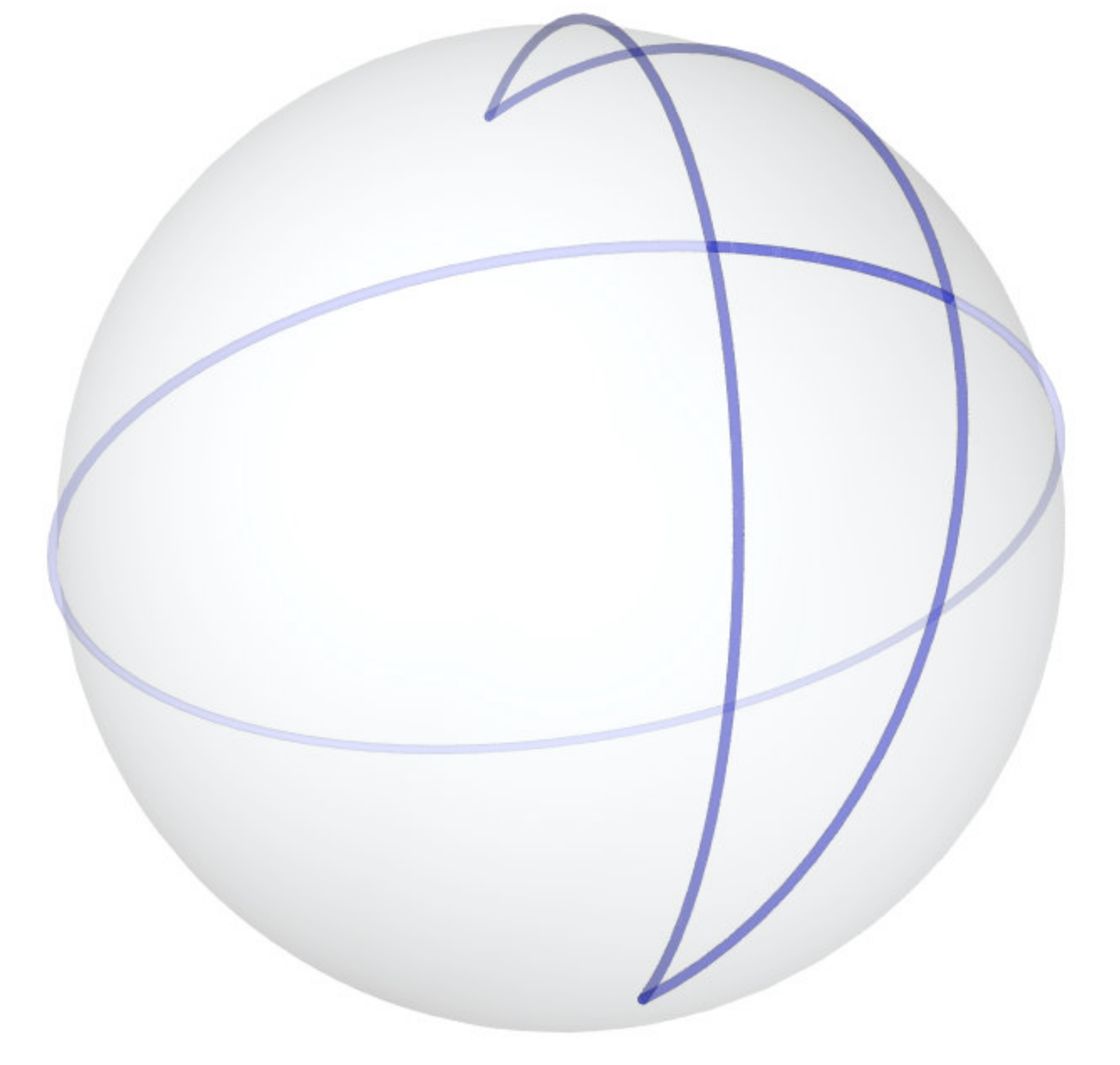} 
    \caption{A spherical digon, or lune. The two antipodal points are connected by two arcs of length $\pi$. The area of the lune equals twice the length of the arc it spans on the great circle orthogonal to both arcs. }
    \figLab{lune}
\end{figure}

The antipodal pair of points where two great-circles cross is dual to a great-circle. This passes through two pairs of antipodal points on the sphere, which are dual to the two original crossing great-circles. These two pairs of points determine four {\em short} arcs on their spanning great-circle, grouped into two antipodal pairs. Two great-circles determine four {\em lunes}, grouped into two antipodal pairs. The duality takes a short arc (and its antipodal) to a lune (and its antipodal). The set of circles crossing an arc (and its antipodal) is mapped, by duality, to a set of points contained in the dual lune, and its antipodal. The duality allows for the definition of a measure on sets of great-circles.

\medskip
\noindent
{\bf Lebesgue measure on the sphere.} The set of all great circles is endowed with Lebesque measure
$\mu$. This is the spherical area of the dual set of antipodal points of the great circles, divided by $2$ to account for the antipodal symmetry. The measure of all the great circles is thus $2\pi$, half of the total area $4 \pi$ of the unit sphere.

\lemmaD{lebesgueMeasure}
{
The Lebesgue measure of the set of great circles crossing an arc of length $\alpha < \pi$ equals $2 \alpha$.
}

\begin{proof}
The dual of the set of circles crossing an arc is a lune of span $\alpha$ and its antipodal. The area of a lune is proportional to the fraction of a great circle spanned by twice its spanned arc, hence $2\alpha$. See Fig.~\ref{fig:lune}.
\end{proof}

\medskip
\noindent
{\bf Belts.} A {\em belt} is the area between two circles at equal distance $\ell$ from a great circle, called its {\em median} equator (or great circle). The width $w=2\ell$ of the belt is the length of the arc orthogonal to its medium equator. The following lemma allows us to measure the set of great-circles contained in a belt. The circles it refers to are arbitrary circles on the sphere, not great-circles. See Fig.~\ref{fig:beltDual}.

\lemmaD{dualBelt}
{
The polar-dual of a belt of width $w$ is a pair of antipodal circles of diameter $w$.  
}

\begin{proof}
Let us denote by $c$ a great-circle and by $p_c$ one of its poles. A great circle $c$ contained in the belt region makes an angle of at most $\frac{w}{2}$ with the belt's equator $e$. Its dual $p_c$ lies at a spherical distance at most this value from the dual $p_e$ of the equator, i.e. it lies on a circle of diamater $w$. 
\end{proof}

\begin{figure}[h]
    \centering
    \includegraphics[width=0.3\textwidth]{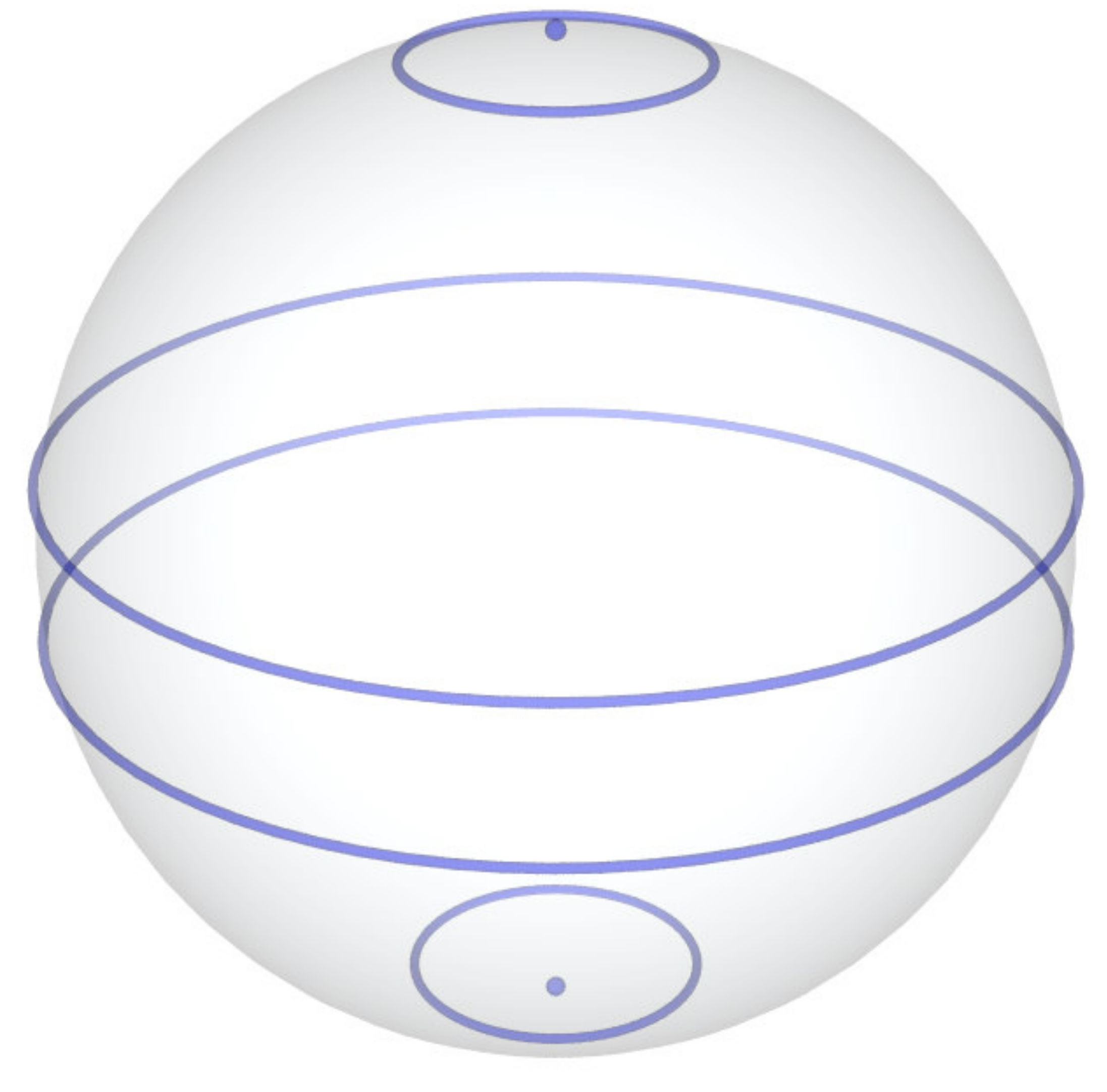}
    \caption{The dual of a belt around an equator is a pair of circles centered at the two poles of the equator.}
    \figLab{beltDual}
\end{figure}

\medskip
\noindent
{\bf Convex spherical polygons.} A spherical (closed) polygon is {\em convex} if for each edge, its spanning great circle contains all the other polygon vertices and edges in one of its two hemispheres. In particular, a lune is a convex spherical polygon. A convex spherical polygon is contained in a hemisphere. Hence the convex polygon has a well-defined 
interior, which is the region lying in the hemisphere.

The following lemma gives a useful estimate of the area of a spherical convex polygon. It is worth noticing that this has no counter-part in the Euclidean setting.

\lemmaD{diameter}
{Let $K \subset S^2$ be a (spherically) convex polygon and $area(K)$ its area. Let $d$ be the maximal diameter  of a circle lying in $K$. Then:
$$area(K) \leq 2d$$
}

\medskip
\begin{proof} 
The statement is true with equality for a lune, whose area is twice the diameter of a maximum inscribed circle, of diameter equal to the arc spanned by the lune.

For an arbitrary polygon, the maximum inscribed circle can be either tangent to two sides, or to more. In the first case, if we eliminate all edges of the polygon not tangent to the circle, we increase the area, and get a lune, for which the statement is valid. In the other case, we eliminate all edges not tangent to the circle, and all but three of those which are tangent. We get a spherical triangle and a circle inscribed in it. Now rotate one of the sides, keeping it tangent to the circle, until it makes a lune with one of the remaining sides, for which the circle is the minimum inscribed one. In the process, the area has increased but the diameter of the circle remained the same.
\end{proof}

\medskip

\sectionD{Main result: flattening medium chains}{algorithm}

\medskip

We describe now an algorithm for planning the motion of a medium-size spherical chain from an arbitrary configuration to one lying on the equator.  We know that a fully expansive global motion may not be possible in this case: indeed, if the endpoints of the chain lie more than $2\pi - \alpha$ apart, where $\alpha$ is the length of the chain, then they must get closer together to reach the final configuration, where they lie at exactly this distance. To design a non-self-intersecting unfolding trajectory, we will patch together segments of expansive trajectories (which are possible when the chain lies in a hemisphere) with other trajectories that are expansive only for subchains.  These subchains will be identified by a {\em separating great circle} cutting through exactly one edge (we will show later that it always exists). Since each of the two parts now lie in separate hemispheres, they can be expanded there, independently. When the unfolding of one of these subchains reaches the boundary of the hemisphere containing it, we recalculate the separating great circle, and continue. Formally:

\bigskip
\noindent
    {\bf Algorithm: Straightening a Spherical Chain of Medium Length}\

	\medskip
    {\bf Input:} A spherical chain $p$ of total length $\alpha$ between $\pi$ and $2\pi$.

	\medskip
    {\bf Output:} A trajectory to unfold the chain onto a great-circle.

	\medskip
    {\bf Method:}

	\medskip
    1. If the chain is straightened onto a great-circle, then stop. 

	\medskip
	2. If the chain lies in a hemisphere, apply an expansive unfolding motion (e.g. Streinu and Whiteley's \cite{streinu:whiteley:origami:2005} adaptation to the hemisphere of the pseudo-triangulation algorithm of \cite{streinu:pseudoTriang:dcg-focs:2005}). Continue for as long as the chain remains hemispherical.

	\medskip
    3. Otherwise, find an edge $e_k$ of the chain such that the measure of all separating great-circles cutting through the edge is the largest. The edge splits the chain into two parts $p'=p[0:k-1]$ and $p''=p[k:n]$.  

\medskip
Fix a belt $b$ of width $w\geq (2\pi - \alpha)/(n+2)$ cutting through the edge $e_k$, as in Fig. ~\figRef{nonHemiChain}. Choose its median as the boundary great-circle (equator) separating the chain into two parts, each lying in its own hemisphere. 
The edge $e_k$ cuts through the equator defining these hemispheres. 

\medskip
In each hemisphere, consider only the part of the edge $e_k$ lying in it as part of the hemispherical subchain.  Pin this edge, and prepare to proceed with the pseudo-triangulation expansive algorithm applied only for one of the subchains, in its hemisphere.

\medskip
	4. Perform the hemispherical expanding unfolding process (as in step 2) for one of them. Stop when either that subchain is straightened or when it hits the middle equator. Then repeat from Step 1. 

\qed

\medskip

We emphasize that Step 2 is carried out only {\em for the part of a chain that lies in a hemisphere}. We remind the reader that the pseudo-triangulation algorithm needs a pinned-down edge. We will use the separating edge for this purpose, or, to be precise, the part of it that lies in the hemisphere where the expansive motion takes place. 

\medskip

The existence of the separating belt of the specified width is proven in Section \secRef{proofs}.

\begin{figure}[h]
	\vspace{-0.15 in}
    \centering
    \includegraphics[width=0.3\textwidth]{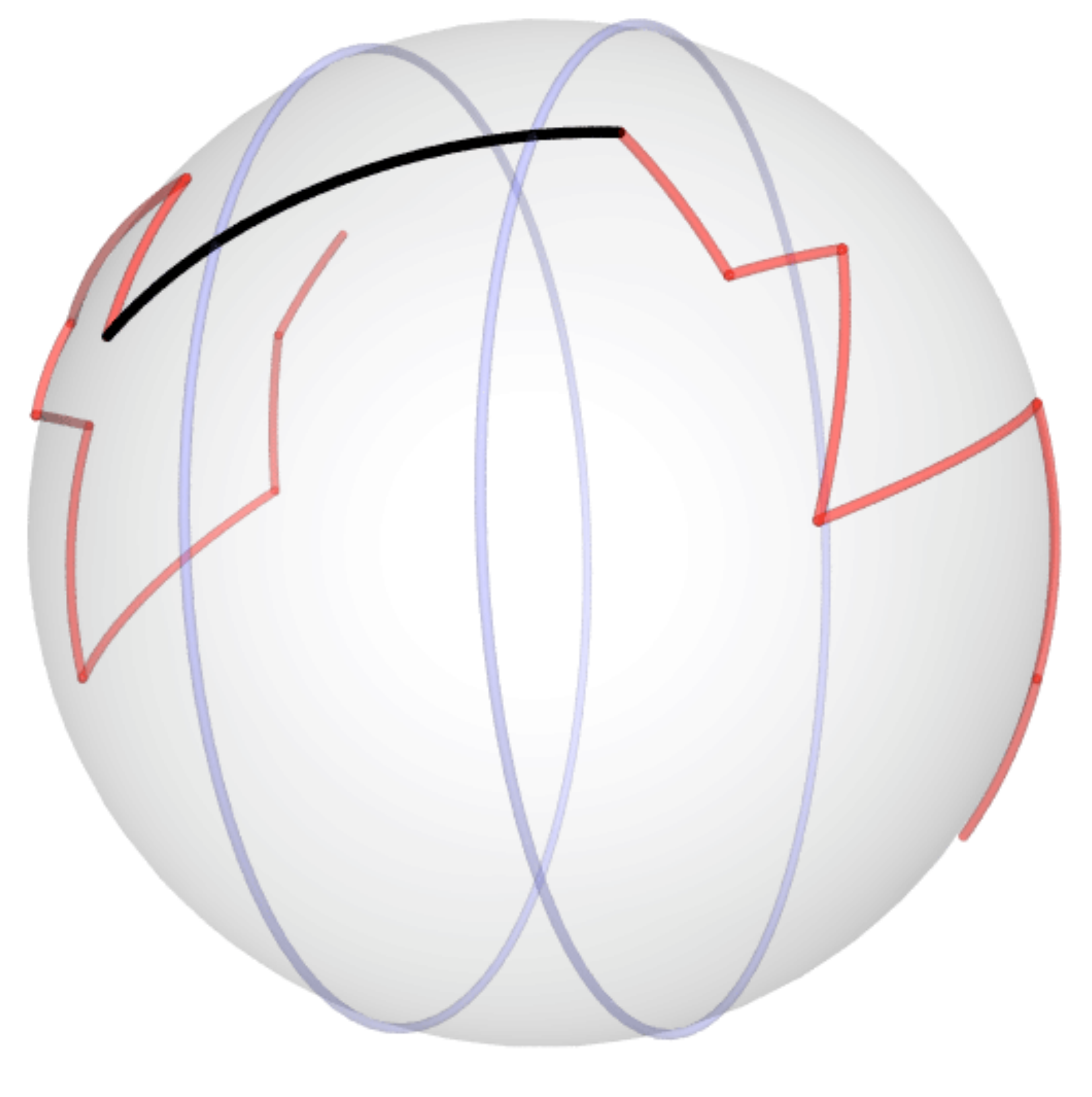}
    \caption{A sphere-spanning chain and a separating belt. The separating edge is shown thicker than the other polygon edges.}
    \figLab{nonHemiChain}
\end{figure}

\medskip
\noindent
{\bf Time analysis.}
The algorithm works in phases, which are expansive on all or part of the chain. A phase is the trajectory between two switches of the separating edge and its belt. We know from \cite{streinu:pseudoTriang:dcg-focs:2005} that each expansive motion lasts at most $O(n^3)$ reconfiguration steps. To complete the time analysis we need to bound the number of phases. 

\medskip

\propD{numberPhases}
{ {\bf (Finiteness of the algorithm)}
A chain $p$ of length $\alpha$ with $n$ edges  will be straightened in
at most $ \frac{2\pi  \alpha}{ (2\pi -\alpha) }(n+2)(n-1)$ phases. 
}

\begin{proof}
Let $\beta_i$ be the small angle between the edge $e_i$ and $e_{i+1}$, for $i=1,\cdots,n-1$.
We will use $\Delta = \Delta(p)$, the sum of these small angles at the inner vertices, as a
measure of progress of the algorithm:

$$\Delta
 =\sum _{i=1}^{n-1} \beta _i$$ 

First, notice that $\Delta$ only grows during the unfolding process. For a straightened chain, it achieves its maximum at $\Delta = (n-1) \pi$. Therefore it suffices to show that during each phase,
$\Delta$ increases with a positive fraction. 

Consider a phase that moves the chain from configuration $p$ to $q$. Define $\delta =\Delta (q)-  \Delta (p)$.

The definition of the phase and the choice for the width of the  belt implies that there exists a vertex $j$ such that its distance $d$ between its original position in $p$ and its final position in $q$ is at least $w/2$, where $w$ is the width of the belt. Indeed, originally the belt was vertex-free, and at the end some vertex hits its middle line.

It is easy to see that  $d \leq  \alpha \delta$. Indeed, $\alpha \delta$ is the distance traveled, on a great circle lying on the sphere, by a point rotating by an angle of $\alpha$ about a center of rotation at distance $\delta$ from it. This certainly exceeds the distance $d$ by which point $j$ was displaced.  Combining these inequalities we obtain that $w/(2\alpha) \leq   \delta$.
Since the belt width was taken to be at least $w \geq \frac{2\pi-\alpha}{n+2}$, the number of phases is at most:
$$ \frac{(n-1)\pi}{\delta} \leq \frac{2\pi\alpha (n-1) }{w} \leq  \frac{2\pi  \alpha }{ (2\pi -\alpha) }(n+2)(n-1)$$

\end{proof}

The rest of the paper contains the proofs.


\medskip

\sectionD{Proofs}{proofs}

\subsectionD{Separating the chain}{separating}

The main technical tool is the existence of a separating great-circle, as needed by Step 2 of the algorithm. More precisely, we need a set of separating circles of large Lebesgue measure.

\thmD{separation}
{{\bf (Separation by great-circle)}
Let $p=(p_0,\cdots,p_n)$ be an open spherical chain of total length between $\pi$ and $2 \pi$. Then, there exists a great circle cutting through at most one edge of the spherical chain.
}

\begin{proof}
Let $p= \{ p_0,\cdots,p_n \}$ be an open spherical chain with $n$ arcs $e_i=(p_{i-1},p_{i})$ of lengths $\alpha_i$, and total length $\alpha:=\sum_{i=1}^n\alpha_i$. 

Consider the set $\cal C$ of all great circles. Its Lebesgue measure is $2\pi$, the area of a hemisphere. 

Let ${\cal C}_i$ be the set of all great circles that cross the edge $e_i$, for $i=1,\cdots,n$. 
Consider the subset $\cal N$ of {\em nice} great-circles that intersect the chain in no more than one point. We partition them into equivalence classes ${\cal N}={\cal N}_0\cup {\cal N}_1\cup \cdots \cup {\cal N}_n$, where ${\cal N}_0$ is the set of circles that do not cross any edge, and ${\cal N}_i$ is the set of those crossing only edge $e_i$. We have ${\cal N}_i={\cal C}_i\cap{\cal N}$, $i=1,\cdots,n$ and ${\cal N}_0=({{\cal C}\setminus{\cup_{i=1}^n{\cal C}_i}})\cap{\cal N}$.

Some of these ${\cal N}_i$ classes may be empty. We want to show that at least one is not empty, and that it has a sizeable measure. Since we use the Lebesque measure of these
sets, which is a measure of area, we can ignore the circles passing through the vertices, which account for lower dimensional subsets.

The theorem now follows from Corollary \corRef{existNiceCircle} below to the following Lemma, which gives a lower bound for the Lebesgue measure $\mu ({\cal N})$.
\end{proof}

\lemmaD{totalMeasure}
{{\bf(Lebesgue measure of nice great circles)}
For any set $e=\{e_1,\cdots,e_n\}$ of $n$ arcs on the sphere, of total length $\alpha$, the Lebesgue measure $\mu ({\cal N}) = \sum_{i=0}^{n}\mu({\cal N}_i)$ of the set of nice great circles $\cal N$ satisfies the inequality:
$$
\eqLab{lowerBound}
\sum_{i=1}^{n}\mu({\cal N}_i) + 2 \mu({\cal N}_0) \geq 2(2\pi -\alpha)
$$
}

Before giving the proof, we observe two straightforward corollaries.

\corD{existNiceCircle}
{{\bf (Nice great circles exist)}
For any set of $n$ arcs on the sphere, of total length $\alpha\leq 2\pi$, either there exists one great circle which doesn't cross any of the arcs, or it crosses exactly one of them. 
}

\corD{largeMeasure}
{{\bf (There exist many nice great circles)}
For any set of $n$ arcs on the sphere, of total length $\alpha < 2\pi$, one of the {\em nice} sets of great circles has large Lebesgue measure $\mu ({\cal N}_i)\geq (4\pi - 2\alpha)/(n+2)$.
}

\medskip
\proofL{ {\lemRef{totalMeasure}.}
We integrate, over the set of all great circles with Lebesgue measure $\mu$, the function $\#(c\cap e)$ giving the number of crossings of a great circle $c$ with the set of arcs $e$. We obtain:

$$
\int_{\cal C}{\#(c\cap e)}{d\mu(c)} = \int_{\cal N}{\#(c\cap e)}{d\mu(c)} + \int_{{\cal C} \setminus {\cal N}}{\#(c\cap e)}{d\mu(c)}
$$

First, notice that over the set $\cal N$, there is at most one crossing. 

$$
c\in {\cal N}_0 \implies  \#(c\cap e) = 0 \implies \int_{{\cal N}_0}{\#(c\cap e)}{d\mu(c)} = 0
$$

$$
c\in {{\cal N}\setminus{\cal N}_0} \implies \#(c\cap e) = 1 \implies 
$$ 
$$
\int_{{\cal N}\setminus{\cal N}_0}{\#(c\cap e)}{d\mu(c)} = \mu({{\cal N}\setminus{\cal N}_0})
$$

Over the rest ${{\cal C} \setminus {\cal N}}$, there are at least two crossings: 

$$
c\in {{\cal C} \setminus {\cal N}} \implies \#(c\cap e) \geq 2 \implies
$$
$$
\int_{{\cal C} \setminus {\cal N}}{\#(c\cap e)}{d\mu(c)} \geq 2 \mu({{\cal C}\setminus{\cal N}}) = 2(2\pi - \mu({\cal N}))
$$

The integral over all great circles is the total length of the set of arcs:

$$
\int_{\cal C}{\#(c\cap e)}{d\mu(c)} = 2 \alpha
$$

\medskip
Putting everything together:

$$
2 \alpha \geq \mu({{\cal N}\setminus{\cal N}_0}) + 2(2\pi - \mu({\cal N}))
$$

\medskip
Finally, using $\mu({{\cal N}\setminus{\cal N}_0})=\sum_{i=1}^{n}\mu({\cal N}_i)$ we get:

$$
\eqLab{lowerBound2}
\sum_{i=1}^{n}\mu({\cal N}_i) + 2 \mu({\cal N}_0) \geq 2(2\pi -\alpha)
$$

Since the sum on the lefthand side has at most $n+2$ non-zero parts, 
there must exist an equivalence class ${\cal N}_i$ whose Lebesgue measure is large: $\mu({\cal N}_i) \geq (4\pi - 2\alpha)/(n+2)$.
\qed
}

\medskip
We have thus shown that for any set of edges, in any placement on the sphere, of total sum strictly less than $2\pi$, there exists a large class of {\em nice} great circles.

The lower bound on the Lebesgue measure of great circles is used next to compute the width of a belt, which in turn was used to bound the number of reconfiguration steps (step 4) in the Algorithm.

\medskip

\subsectionD{Wide belts}{wideBelt}

\medskip

Finally, we give the bound on the width of a belt chosen to mark the progress made during a phase of the algorithm.

\medskip
We start with a straightforward observation. Recall that $\beta_i$ is the small angle between the edge $e_i$ and $e_{i+1}$, for $i=1,\cdots,n-1$.

\propD{notDecrease}
{During this unfolding process, the angles $\beta_i$ never decrease.
}

\begin{proof}
	Indeed, the motion is expansive within one hemisphere, so all the angles inside a hemisphere increase. And all the vertices lie inside one hemisphere or the other.
\end{proof}

We are ready to bound the belt width. Let $p$ be an arbitrary chain configuration.
There are two cases to consider: (a) if the chain is not contained in a hemisphere, or (b) if it is.
In the first case (a), from Lemma ~\lemRef{totalMeasure} we obtain: $\mu({\cal N}_0)=0$ and at least one of the classes ${\cal N}_i$ is large, $\mu({\cal N}_i)\geq {{4\pi-2\alpha}\over{n}}$. But we do not need to be so precise. In either case,
we can safely bound $\mu({\cal N}_i)\geq {{4\pi-2\alpha}\over{n+2}}$.  

\lemmaD{beltBound}{
{\bf (Belt bounds)}
\begin{enumerate}
    \item If $\mathcal{N}_0$ is a large equivalence class, then there exists a
    belt of width at least $w \geq (2\pi - \alpha)/(n+2)$
    which does not intersect the chain $r$.
    \item If $\mathcal{N}_i, i \neq 0$ is a large equivalence class, then
    there exists a belt of width $w \geq (2\pi - \alpha)/(n+2)$ which
    does not intersect the chain except for the edge $e_i$.
\end{enumerate}
}

\begin{proof}
The two cases are similar, so we do only one. Define the set of points dual to the great circles in ${\cal N}_i$ as
${\cal N}_i^* = \{c^*: c \in {\cal N}_i\}$.  It is easy to see that it is a
convex spherical polygon. Its area is at least $(4\pi - 2\alpha)/(n+2)$. Therefore, applying Lemma \lemRef{diameter}, it follows that it contains a circle  $C$ of diameter $w=(2\pi - \alpha)/(n+2)$. Then the
great-circles dual to points in $C$ sweep a belt $C^* \subset {\cal N}_i$ of
width $w$.
\end{proof}

This completes the proof.


\medskip

\sectionD{Concluding remarks and open questions}{conclusions}

\medskip

We have shown that there exists a non-colliding motion that unfolds a spherical open chain of length less than $2\pi$, thereby settling the question of reconfigurability and unfoldability of single-vertex origamis. We conclude with some remaining open questions.

There is an asymmetry in the usage of expansive motions in a hemisphere, where the number of pseudo-triangulation induced steps can be bound by $O(n^3)$, a function of $n$, and the number of belt-shrinking steps, for which the bound depends on the length deficit $2\pi-\alpha$ of the chain. The main remaining question for medium-length open {\em single-vertex origamis} is: 

\medskip
\begin{open}
	Design an algorithm based on motions induced by one-degree-of-freedom mechanisms on the sphere (which will necessarily be partially expansive, partially contractive) for which the number of events (when the mechanism's bars align or encounter some similarly easy-to-verify events) can be bounded {\em only} in terms of the number $n$ of bars of the chain.
\end{open}

An algorithmic question remains to be investigated for long open and closed {\em single-vertex origamis}:

\begin{open}
	Decide, algorithmically, whether a single vertex-origami whose length exceeds $2\pi$ can be reconfigured, without collisions, between two given configurations.
\end{open}

\medskip
As a consequence of our work, the study of general {\em non-colliding} origami folding can now focus on the interaction between panels incident with distinct vertices of the origami pattern, since locally, each single-vertex sub-unit has non-colliding motions. This appears to be a difficult topic, whose systematic investigation is yet to be undertaken.

\medskip
\noindent
{\bf Acknowledgement.}
We thank Yang Li for help with the preparation of the figures. 

Partial funding for the authors was provided by an NSF International Collaboration grant.
The second author was also supported by NSF CCF-0728783 and by a DARPA {\em Mathematical Challenges} grant, under {\em Algorithmic Origami and Biology}. All statements, findings or conclusions contained in this publication are those of the authors and do not necessarily reflect the position or policy of the Government.

\end{document}